\documentstyle[11pt,epsfig,epsf]{article} 
\def\baselinestretch{1.3}
\newcommand{\ba}{\begin{array}}
\newcommand{\ea}{\end{array}}
\newcommand{\bd}{\begin{displaymath}}
\newcommand{\ed}{\end{displaymath}}
\newcommand{\be}{\begin{equation}}
\newcommand{\ee}{\end{equation}}
\newcommand{\bea}{\begin{eqnarray}}
\newcommand{\eea}{\end{eqnarray}}

\begin{document}
\begin{flushright}
{\large MRI-P-010801\\ 
hep-ph/0108150}\\
\end{flushright}

\begin{center}
{\Large\bf  Effects of lepton number violating interactions on $t \bar t$ 
production at NLC
}\\[20mm]
{\sf Anindya Datta} \footnote{E-mail: anindya@mri.ernet.in}, 
\\
{\em Harish-Chandra Research Institute,\\
Chhatnag Road, Jhusi, Allahabad - 211 019, India}
\end{center}

\vskip 20pt
\begin{abstract}
  
  We discuss the effects of lepton number violating interactions
  namely, R-parity violation and leptoquarks on top-quark pair
  production at the upcoming $e^+ e^-$ linear colliders. Effects of
  $SU(2)$ singlet, doublet and triplet leptoquark interactions are
  investigated.  R-parity violating minimal supersymmetric standard
  model also allows certain kinds of lepton number violating
  interactions which are same as singlet leptoquarks with left-handed
  interactions. We have calculated the cross-section of $e^+ e^-
  \rightarrow t \bar t$ in presence of the above interactions. With 
  conservative values of lepton number violating coupling strengths
  we got enhancement of top-pair production cross-section in all of
  the above cases. 
\end{abstract}

\vskip 1 true cm

\setcounter{footnote}{0}

\def\baselinestretch{1.8}

\section{Introduction}
It is well known that lepton-number conservation in
the Standard Model (SM) is an accidental symmetry. It is a mere
outcome of particle content and gauge structure of the SM. In many
extensions of the SM, lepton number violating interactions occur in a
natural way. Minimal Supersymmetric Standard Model (MSSM) without
R-parity \cite{R_parity} and non-SUSY theories with leptoquarks
\cite{lepto_quark} are well cited examples of it. The key feature of
these theories, relevant for the following analysis, is the presence,
in their spectrum, of a scalar (leptoquark) which couples to a quark
and a lepton at the same time.  Leptoquarks arise in many models of
extended gauge symmetry including the grand unified theories.  In many
of these models, vector leptoquarks can also arise.  The gauged
vector leptoquarks are superheavy. Their mass is related to the scale
of spontaneous breaking of the lepton number. On the other hand,
interactions involving the non-gauged vector leptoquarks are
non-renormalizable.  Several interesting phenomenological analyses
have been done considering both of these interactions. In this article,
we will focus on how these scalar leptoquark (a class that also includes
the squarks in R-parity violating SUSY) interactions can modify the
top-quark pair production cross-section significantly at the next
generation $e^+ e^-$ colliders.  The choice of this particular process
has several advantages. The foremost is the copious production of top
pairs at these machines.  Also the cleaner environment of leptonic
colliders helps one to make the precision studies like measuring such
deviations more accurately, if exists, from the SM expectation. One of the
major goals of these $e^+ e^-$ machines is to measure the top-quark
interactions to a high level of precision \cite{nlc}.  Measurement of
the lepton number violating couplings involving light quarks (mainly
of first generation) and leptons can also be done at the hadron
colliders by studying the processes like Drell-Yan pair production of
the leptons.  But at a hadronic machine, the couplings in which we will
be interested in our analysis can only be probed in the decays of the
heavy quarks. Though production cross-section of a heavy quark pair is
huge at a hadron collider, presence of competing QCD backgrounds may
intervene such precise measurements.

Baryon (B) and lepton number (L) violating processes involving the
top-quark have been investigated by several authors. For example,
effects of B- and L-violation in top-quark production at hadronic
colliders have been analysed in ref. \cite{dkg} in the context of
R-parity violating SUSY. People have extensively studied the single
top production \cite{single_top} and decay of top-quark
\cite{top_decay} mediated by R-parity violating interactions.  Effects
of R-parity violation on the top mass have been discussed in
\cite{top_mass}. So a lot of attention has already been given to the
top phenomenology \cite{top_pheno} in the context of R-parity
violation. Though leptoquark interactions has similarities with that
of R-parity violating SUSY, in some of the cases the chiral structure
of the relevant couplings differs from it. People also have payed a
lot of attention to leptoquark phenomenology. Apart from direct
leptoquark searches at future lepton and hadron colliders
\cite{lepto_search}, effects of these interactions have been studied
in the context of neutrinoless double beta decay \cite{klapdor}, muon
anomalous magnetic moment \cite{anomal} and needless to mention, to
explain the HERA anomaly \cite{hera_ano}. Indirect effects of
leptoquark interactions have also been investigated in the context of
$e^+ e^-$, $e \gamma$ and hadronic colliders \cite{indirect_lepto}. In
this paper we will try to discuss, in some detail, how these lepton
number violating couplings can affect the pair production and 
decay of the heaviest quark.  This has been studied previously in
\cite{lepto_debchou} in a slightly different manner.  Using polarised
electron beam in $e^+ e^-$ collision, constrains are derived, in the
above reference, on leptoquark mass and couplings by comparing (and
then doing a $\chi^2$-analysis) the angular distribution of leptoquark
mediated process with that of the pure SM. It was
shown in this ref. that an 1 TeV, $e^+ e^-$ collider will be more
efficient than a 500 GeV machine, in exploring/excluding the
parameter space of leptoquark interactions. We will focus this point
more on later.  People have also considered the effects of
vector leptoquarks on $t \bar t$ production from $e^+ e^-$ collision
\cite{lepto_aliev}. Authors in ref. \cite{lepto_aliev} also used
polarised $e^-$ beams to differentiate the vector leptoquark
interactions from the SM. They have presented the variation of total
number of $t \bar t$ events with vector leptoquark mass assuming the
leptoquark couplings to an $e$ and $t$ of the order of unity. Though
the structures of the vector leptoquark interactions are different
from those of the scalar leptoquarks, qualitatively the variation of
production cross-section with leptoquark mass, agrees with us.  In
this article, we will concentrate on how the total cross-section would
change in presence of such particles and how angular asymmetry in $t
\bar t$ production and decay can be used to discriminate the different
types of leptoquark interactions.  Plan of the rest of the article is
as follows. In the next section, we will discuss the models briefly
with special emphasis on the relevant couplings and the similarities
and differences in two models of our interest. The third section will
contain the result of our analysis followed by a conclusion in the
last section.

\section{Relevant interactions}
In this section we will discuss briefly the phenomenology of lepton
number violating interactions in the context of $t \bar t$ production
in $e^+ e^-$ collision. As we emphasised earlier, two main kinds of
models which allow these  interactions are MSSM with R-parity
violation and non-SUSY theories with leptoquarks. As it has been
noted in the literature, unless a discrete symmetry \footnote{This
  symmetry is called R-symmetry.  R is defined as: $(-1)^{3(B-L)
    -2S}$.  All the SM fields have $R = 1$ and all the SUSY partners
  have $R = -1$. Apart from ruling out both $B$ and $L$ violating
  interactions this symmetry has an additional consequence of
  rendering the lightest super-particle absolutely stable.}  is
introduced by hand, the MSSM superpotential contains the following 
terms \cite{rp_lag}: 
\be 
W_{\not R} = \lambda_{ijk} {\hat
  L}_i {\hat L}_j {\hat E}_k^c + \lambda_{ijk}' {\hat L}_i {\hat Q}_j
{\hat D}_k^c + \lambda_{ijk}''{\hat U}_i^c {\hat D}_j^c {\hat D}_k^c +
\epsilon_i {\hat L}_i {\hat H}_2 
\label{super_pot}
\ee 
However, such a symmetry is {\em ad hoc}. So it is of interest to
consider possible violation of this symmetry especially when it has
some interesting experimental consequences in detecting the
supersymmetric particles \cite{exp_rp}.  One can easily see that the
first two and the last term in the superpotential violate the lepton
number/flavour explicitly while the third term breaks the baryon
number. As we are interested in the top pair production in electron
positron annihilation, we will be interested in the second term. One
can expand this piece in terms of the normal fields. This in turn,
yields (with many other) the relevant interactions involving a lepton
and a quark along with a squark. One can easily write the 
interaction of our interest.

\be
{\cal L}_{\lambda'}  = 
  -\lambda'_{13k}\;({\tilde d}^k_R)^\ast ({\bar e}_L)^c t_L + h.c.
\label{rp_vio}
\ee

Now we will turn our attention to the leptoquark interactions. The 
interactions necessary for our purpose are listed in a tabular form in
the following \cite{lepto_lag}.
\begin{table}[h]
\begin{center}
\begin{tabular}{|c|c|c|}
\hline
& &  \\[-2ex]
 leptoquark Type  &  Coupling & $SU(3)_c\times SU(2)_L\times U(1)_Y$\\
& &  \\[-2ex]
 \hline
 \hline
& &  \\[-2ex]
 $\Phi_1$ & $\left[\lambda_{ij}^{(1)} {\bar Q}_{L j}^c L_{L i} 
 + {\tilde \lambda}_{ij}^{(1)} {\bar u}_{R j}^c e_{Ri}\right]\Phi_1
 $ & 
        (${\bar 3}$, $1$, ${2\over 3}$)\\ 
& &  \\[-2ex]
 \hline
& &  \\[-2ex]
 $\Phi_2$ & $\left[\lambda_{ij}^{(2)} {\bar Q}_{L j} e_{R i} 
 + {\tilde \lambda}_{ij}^{(2)} {\bar u}_{R j} L_{Li}\right]\Phi_2
 $ & 
        ($3$, $2$, ${7\over 3}$)\\ 

& &  \\[-2ex]
 \hline
& &  \\[-2ex]
 $\Phi_3$ & $\lambda_{ij}^{(3)} {\bar Q}_{L j}^c L_{L i} \Phi_3$ & 
        (${\bar 3}$, $3$, ${2\over 3}$)\\ 

 \hline
\end{tabular}
\end{center}
\label{table1}
\caption{{\it Different kinds of leptoquark interactions relevant for our analysis.
R-parity violating MSSM interaction in eqn.\ref{rp_vio} corresponds to 
the left-handed (proportional to $\lambda ^{(1)}_{13}$) interaction of $\Phi_1$.}}
\end{table}

Here we have suppressed the $SU(2)$ indices. One can very easily write
the interactions relevant for our purpose involving $e$, $t$ and a
particular leptoquark from the above table. Below we write the
interaction Lagrangians separately for singlet, doublet and triplet
leptoquarks \footnote{In eqn. \ref{lepto} and Table. \ref{table1},
  one should not confuse the $\lambda$ couplings with that in
  eqn.\ref{rp_vio}. The $\lambda$ couplings here, have more
  similarities with the $\lambda ^{\prime}$-coupling in eqn.
  \ref{super_pot}.}
\bea
{\cal L}_1 &=& -\left[ \lambda^{(1)}_{13}\; ({\bar e})^c \,P_L \,t
 + {\tilde\lambda}^{(1)}_{13}\; ({\bar e})^c \,P_R \,t \right] \phi_1+ 
 h.c. \nonumber \\
{\cal L}_2 &=& \left[ \lambda^{(2)}_{13}\; {\bar t} \,P_L \,e
 - {\tilde\lambda}^{(2)}_{13}\;{\bar t} \,P_R \,e \right] \phi_2+ 
 h.c. \nonumber \\
{\cal L}_3 &=&  \lambda^{(3)}_{13}\;({\bar e})^c \,P_L \,t
  \,\phi_3+  h.c.
\label{lepto}
\eea

There are some similarities and differences between the above
interactions and that in eqn. \ref{rp_vio}. The triplet and the
left-handed singlet (proportional to $\lambda^{(1)}_{13}$) have
similar structures to the R-parity violating interaction. Charges of
the leptoquarks in such cases are also the same with that of the
squark involved in eqn.\ref{rp_vio}.  At the same time $\phi_1$ has a
coupling with $e$ and $t$ which is right-handed in nature. This type
of interaction is not allowed in SUSY. $SU(2)$ doublet leptoquark
$\phi_2$ has a similar kind of interactions like $\phi_1$. The only
difference is it's electromagnetic charge which is equal to
$\frac{5}{3}$.

The operators, those will contribute to the top-quark pair production via
$e^+e^-$ annihilation, follow very easily from the Lagrangian. They are
given in Table 2.
\begin{table}{h}
\begin{center}
\begin{tabular}{|c|c|c|}
\hline
1. & Squark, & $|\lambda^{(i)}_{13}|^2 \left( {\bar t} P_R e^c\;\;
{\bar e}^c P_L t  
\right)
 \widehat{\phi_i \phi_i^\ast}$ \\
&Singlet/Triplet- & \\
&leptoquark (Left-handed)& ({\bf RL}) \\
\hline
2. & Singlet- &$|\tilde \lambda^{(i)}_{13}|^2 \left({\bar t} P_L e^c 
\,\,{\bar e}^c P_R t  \right) \widehat{\phi_i \phi^\ast_i}$ \\
&leptoquark (Right-handed)& ({\bf LR}) \\
\hline
3. & Singlet- &$|\lambda^{(1)}_{13} \tilde \lambda^{(1)}_{13}| 
\left( {\bar e}^c P_\alpha t \,
\,{\bar t} P_\alpha e^c \right) \widehat{\phi_1 \phi^\ast_1}$ \\
&leptoquark (Right-Left)& $\alpha = L,R$ ({\bf LL, RR}) \\ 
\hline
4. & Doublet- &$|\lambda^{(2)}_{13}|^2 \left( {\bar t} P_L e \,
\,{\bar e} P_R t \right) \widehat{\phi_2 \phi_2^\ast}$ \\
&leptoquark (Left)& ({\bf LR})\\
\hline
5. & Doublet- &$|\tilde \lambda^{(2)}_{13}|^2 \left( {\bar t} P_R e \,
\,{\bar e} P_L t \right) \widehat{\phi_2 \phi_2^\ast}$ \\
&leptoquark (Right)& ({\bf RL})\\
\hline
6. & Doublet- &$|\lambda^{(2)}_{13} \tilde \lambda^{(2)}_{13}| 
\left( {\bar t} P_\alpha e \,
\,{\bar e} P_\alpha t \right) \widehat{\phi_2 \phi_2^\ast}$ \\
&leptoquark (Right-Left)& $\alpha = L,R$ ({\bf LL, RR}) \\
\hline
\end{tabular}
\end{center}
\label{table2}
\caption{{\it Different types of operators contributing to  the process
$e^+ e^- \rightarrow t \bar t$, made out of interactions in eqns. 
\ref{rp_vio},\ref{lepto}. R-parity violating MSSM corresponds to case 1.
For the first two cases $i$ can be 1 or 3.}}
\end{table}

\begin{figure}[htb]
\centerline{
\epsfxsize=7.cm\epsfysize=4cm
                     \epsfbox{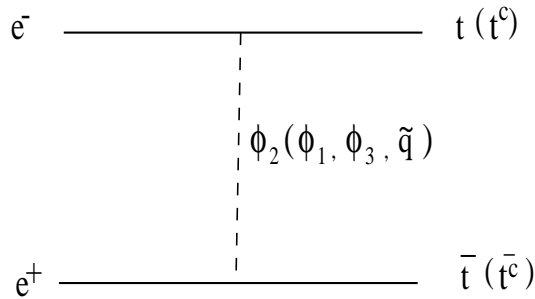}
}
\caption{\em Feynman diagram for the process $e^+ e^- \rightarrow t \bar t$
in $R_p$ violating SUSY or leptoquarks. }

\label{fig:prod}
\end{figure}

Apart from SM s-channel diagram (mediated by $\gamma$ or $Z$), one has
to calculate an extra diagram mediated by the squark or leptoquarks
(see fig.\ref{fig:prod}) due to these lepton-number violating
interactions. Looking at the Lagrangians, one can easily check that
in R-parity violating contribution, one vertex is proportional to
$P_L$ and the other is proportional to $P_R$. While in the leptoquark
mediated contributions $P_L$ or $P_R$ can arise in both the vertices.

For the sake of completeness, we write down the expressions for the 
amplitudes, arising due to different types of interactions listed in
Table. \ref{table1}, along with the SM. 

\bea
{\cal M}_{SM}&=& -\;\frac{1}{s - m_V^2 + i\;m_V\;\Gamma_V}\;
\left(\bar v(p_1)\;\gamma_\mu\;(a_e + b_e \gamma_5)\;u(p_2)\right)\;
\left(\bar u(p_3)\;\gamma^\mu\;(a_t + b_t \gamma_5)\;v(p_4)\right)
 \nonumber \\
{\cal M}^{S/T}_{LQ}&=&\frac{|\lambda|^2,|\tilde \lambda|^2, 
\lambda\tilde\lambda}{t - m_{\phi}^2}\; \left(\bar u(p_3)\;P_i\;u(p_1)\right)\;
\left(\bar v(p_2)\;P_j\;v(p_4)\right) \nonumber \\
{\cal M}^{D}_{LQ}&=&\frac{|\lambda|^2,|\tilde \lambda|^2, 
\lambda\tilde\lambda}{t - m_{\phi}^2}\;\left(\bar u(p_3)\;P_i\;u(p_2)\right)\;
\left(\bar v(p_1)\;P_j\;v(p_4)\right) \nonumber \\
\eea

The first (${\cal M}_{SM}$) of the above equations stands for the two
SM s-channel diagrams. For the photon- exchange diagram, $b_e = b_t =
0$, $a_e = -e, a_t = \frac{2}{3}e$ and $m_V = \Gamma_V = 0$. For the
$Z$-exchange diagram, $a_e = \frac{g}{\cos\theta_W}\;(-\frac{1}{4} +
\sin^2\theta_W)$, $b_e = \frac{g}{4 \cos\theta_W}$, $a_t =
\frac{g}{\cos\theta_W}\;(\frac{1}{4} - \frac{2}{3} \sin^2\theta_W)$,
$b_t = -\frac{g}{4 \cos\theta_W}$. Next two expressions, ${\cal
M}^{S/T}_{LQ}$ and ${\cal M}^{D}_{LQ}$, are for singlet/triplet and
doublet leptoquark mediated diagrams respectively. $p_1$, $p_2$, $p_3$
and $p_4$ are the momenta of $e^+$, $e^-$, $t$ and $\bar t$.  The
Mandelstum variables are defined as: $s = (p_1 + p_2)^2$ and $t = (p_1
- p_3)^2$ for singlet/triplet and $t = (p_2 - p_3)^2$ for doublet
leptoquarks. Amplitudes for leptoquark mediated diagrams, are
proportional to $|\lambda|^2$ when $P_i = P_R$, $P_j = P_L$; to
$|\tilde \lambda|^2$ when $P_i = P_L$ and $P_j = P_R$ and to $\tilde
\lambda \lambda$ when both are $P_L$ or $P_R$.  Following the tables 1
and 2, triplet leptoquark contribution can only be proportional to
$|\lambda|^2$. The other cases do not arise for the triplet leptoquark
mediation.

Now let us discuss the experimental bounds on the relevant couplings.
The R-parity violating contribution is proportional to the coupling
$\lambda '_{13k}$, where $k$ is the generation index. We will
consider only one R-parity violating coupling to be non-zero at a
time. Looking at the literature \cite{rp_bounds}, one can check easily
that the coupling $\lambda '_{132}$ is the most loosely constrained
\cite{rp_lag}\footnote{This particular coupling is constrained from
  the forward-backward asymmetry in $e^+ e^-$ collision.}. So we will
use this particular coupling in the following analysis. This implies
the exchanged squark in  fig.\ref{fig:prod} is the supersymmetric
partner of s-quark. The same constrains would also exactly apply on
the left-handed singlet $(\lambda_{13}^{(1)})$ and triplet
($\lambda^{(13)}_3$) leptoquark couplings to $e$ and $t$.  The
product of the couplings $\lambda ^{(i)}_{13} {\tilde \lambda^{(i)}_{13}}$
in these two cases are unconstrained. The $SU(2)$ doublet leptoquark
couplings $\lambda^{(2)}_{13}$, ${\tilde \lambda}^{(2)}_{13}$ (left and
right) are individually constrained from the $e^+e^-$ partial decay
width of $Z$-boson \cite{gg_lepto}. It is interesting to observe that
the left-handed couplings are more stringently constrained than their
right-handed counterparts. Numerical values of the upper bounds on the
left-handed couplings of the $SU(2)$ doublet leptoquarks are
comparable with the upper bounds obtained for $R$-parity violating
coupling strengths. There is no upper bound on the product of the left- and
right-handed leptoquark couplings. So we may take their values as
free parameter, keeping in mind that the value  should be perturbatively
viable.

\section{Discussion of the results}

We will discuss, in this section, the numerical results from our
analysis.  We have only estimated the Born level diagrams
corresponding to the operators in Table 2. All the coupling constants
scale with the scalar mass. In the case of R-parity violation,
$\lambda^{\prime}_{132}$ scales linearly with $\tilde t_L$ mass.  This
particular coupling is constrained to be less than $0.28$ for a $100
~GeV$ $\tilde t_L$ mass \cite{rp_bounds}. As we discussed earlier,
this bound equally applies to the left-handed singlet and triplet
leptoquark couplings.  We will also use the same values for $\tilde
\lambda^{(1)}_{13}$ ($= 0.3$) and the product $\tilde
\lambda^{(1)}_{13} \lambda^{(1)}_{13} \;(=0.09)$ as there are no
phenomenological bounds available for those.  Again for numerical
values of the couplings involving doublet leptoquarks we follow the
ref.  \cite{gg_lepto}.  For a $100 ~GeV$ scalar, upper bound for left-
($\lambda^{(2)}_{13}$) or right-type ($\tilde \lambda^{(2)}_{13}$)
coupling is almost the same (and is nearly equal to 0.4). While the
upper bound on the $\lambda^{(2)}_{13}$ coupling is not very sensitive
to leptoquark mass, upper bound on the other one rises pretty fast
with the scalar mass. We will use the same values as before (like
singlet and triplet leptoquark) for these couplings which makes our
estimate conservative.

In figure. 2 and 3 we present the numerical estimates of the
cross-sections.  We do not consider any higher order corrections to
the process of our interest. Higher order corrections are important
\cite{nlo}.  In the case of SM, inclusion of higher order effects
increase the cross-section significantly. The aim of this paper is to
show the enhancement of the total cross-section (of $t \bar t$
production) over its SM value, when one includes the lepton number
violating interactions arising from leptoquarks or R-parity violation.
We have calculated the cross-section at centre-of-mass energies away
from the $t \bar t$ threshold. Around the centre-of-mass energy of 350
$GeV$ ($\sim 2\,m_t$), threshold effects are very important
\cite{threshold}. And we wanted to avoid this extra complicacy. But
this does not reduce the very essence of our analysis.

\begin{figure}
\centerline{
\epsfxsize= 7 cm\epsfysize=7.0cm
                     \epsfbox{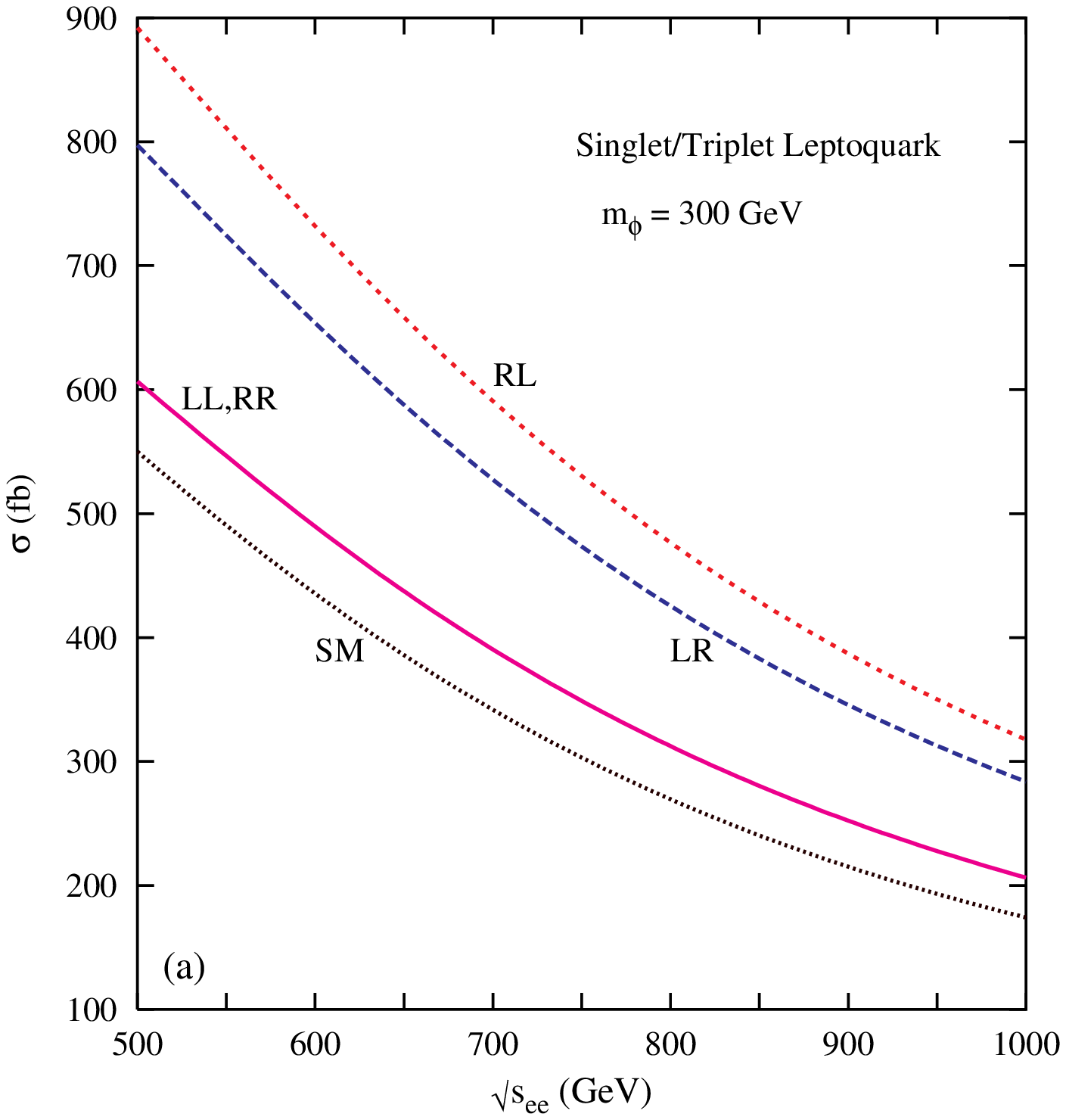}
\epsfxsize=7.0 cm\epsfysize=7.0cm
                     \epsfbox{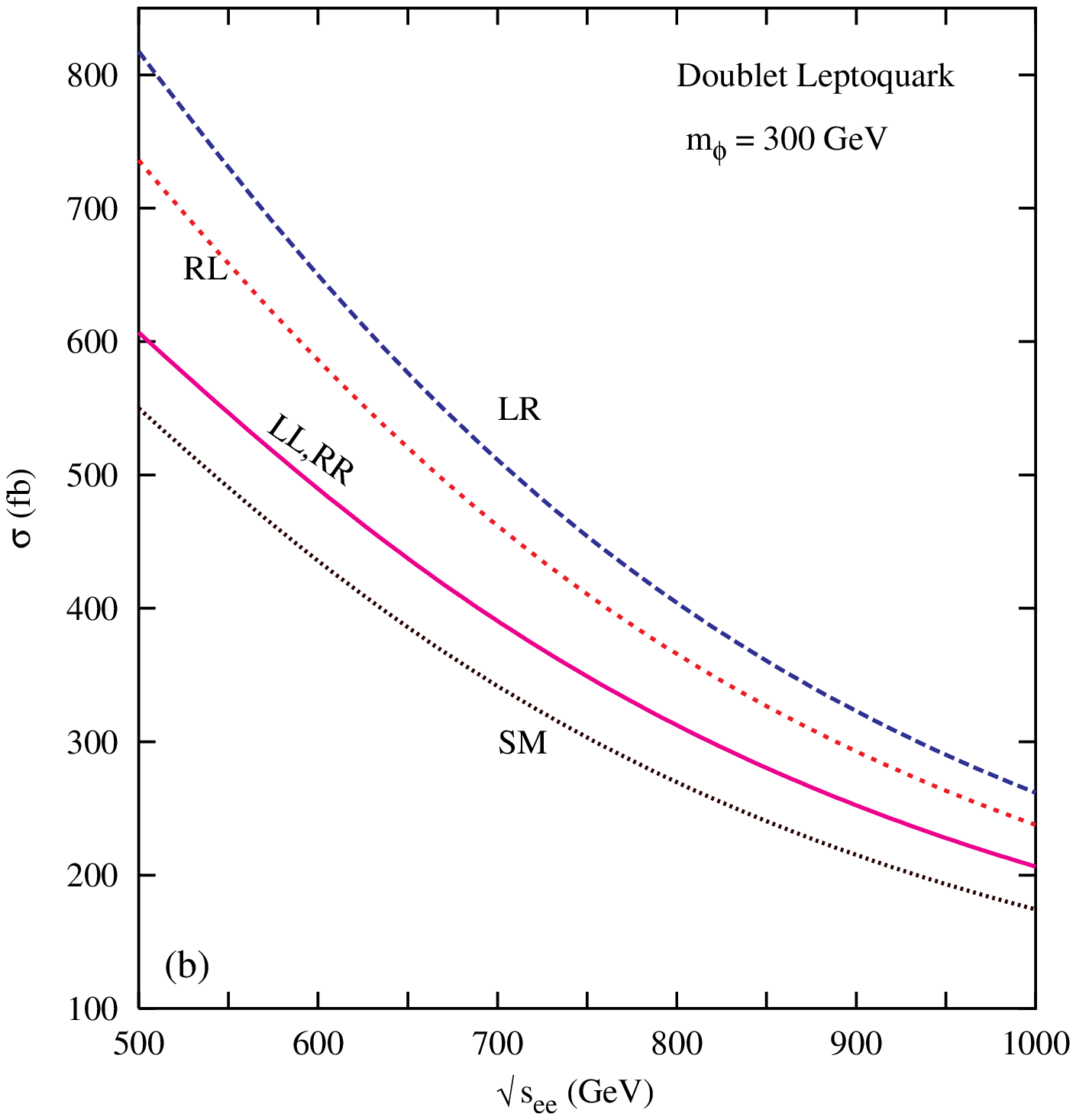}
}

\caption{{\it Variation of top-pair production cross-section (in presence of (a) singlet,
    triplet leptoquarks, R-parity violating interactions and (b) doublet
   leptoquarks)
    with $e^+ e^-$ centre-of-mass energy, $\sqrt{s_{ee}}$. Leptoquark
    mass ($m_\phi$) is fixed at 300 GeV.  For comparison we have also
    plotted the pure SM contribution. Different lines are for different 
 kinds of interactions. Legends follow from the Table 2. The curve 
 marked by $RL$ in (a) corresponds to $R$-parity violating SUSY. 
    }}

\end{figure}

In figure 2a., we plotted variation of the total cross-section of top
pair production with centre-of-mass energy for singlet and triplet
leptoquarks.  For the purpose of illustration, we present the
cross-section with one value of scalar mass (say 300 GeV, which is
well above the bounds quoted by CDF and D0 \cite{cdf_d0_bound} from
Tevatron search limits for squarks and leptoquarks.) and setting the
value of all ($\lambda_i, \tilde \lambda_i, i = 1,2,3$) the couplings
at say $0.3$.  There are several cases of interest, following Table 2.
The $LL$ and $RR$ types of interactions do not interfere with the SM
contribution. It is also worth mentioning here that $LL, RR$ and $LR$
lines in fig.2a come from the singlet leptoquarks  only.  The others,
namely $LR$ and $RL$, interfere constructively with the SM. For
comparison, we plotted the pure SM contribution as well. It is clear
from the figures that presence of any one kind of lepton number
violating interactions increase the $t \bar t$ cross-section over its
SM value. It is worth mentioning that the R-parity violating MSSM
contribution corresponds to the $RL$ case of figure 2a.  Incidentally,
this case shows the maximum enhancement.  MSSM with or without
R-parity conservation is one of the strongest contender of physics
beyond the SM which we expect to see at the next generation of
colliding machines. So any enhancement of top cross-section at
$e^+e^-$ linear colliders may be a positive signal of this kind of
scenario.  The $LR$ case is also interesting to observe.  Here also
the enhancement is pretty prominent. Finally the $LL$ or $RR$, which
can only arise from leptoquark interactions (this is also true for
$LR$ case), enhances the total cross-section by 10$\%$ or so over the
entire range of centre-of-mass energy we have considered.

Plots in Fig. 2(b) are for doublet leptoquarks. Structure of the
interactions, here, are little different from that of the singlet
case.  Otherwise one easily see, comparing the figures 2(a) and (b)
that contributions are nearly the same for both the cases. Here again
the LL or RR type of interactions do not interfere with the SM.
Enhancement of $t \bar t$ cross-section is also exactly the same in
magnitude as the singlet case with LL or RR interaction.

\begin{figure}[ht]

\centerline{
\epsfxsize= 7.0 cm\epsfysize=7.0cm
                     \epsfbox{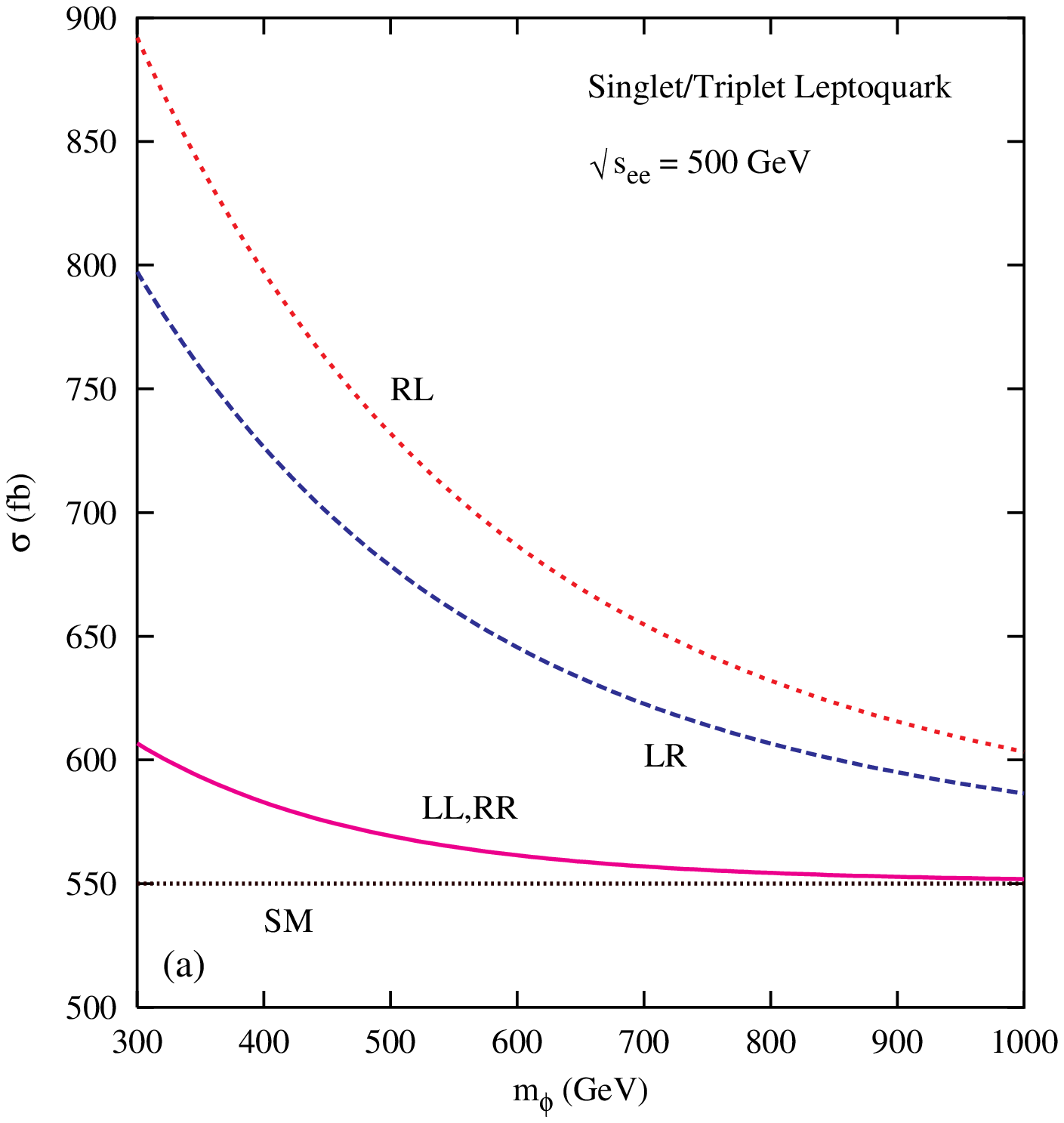}
\epsfxsize=7.0 cm\epsfysize=7.0cm
                     \epsfbox{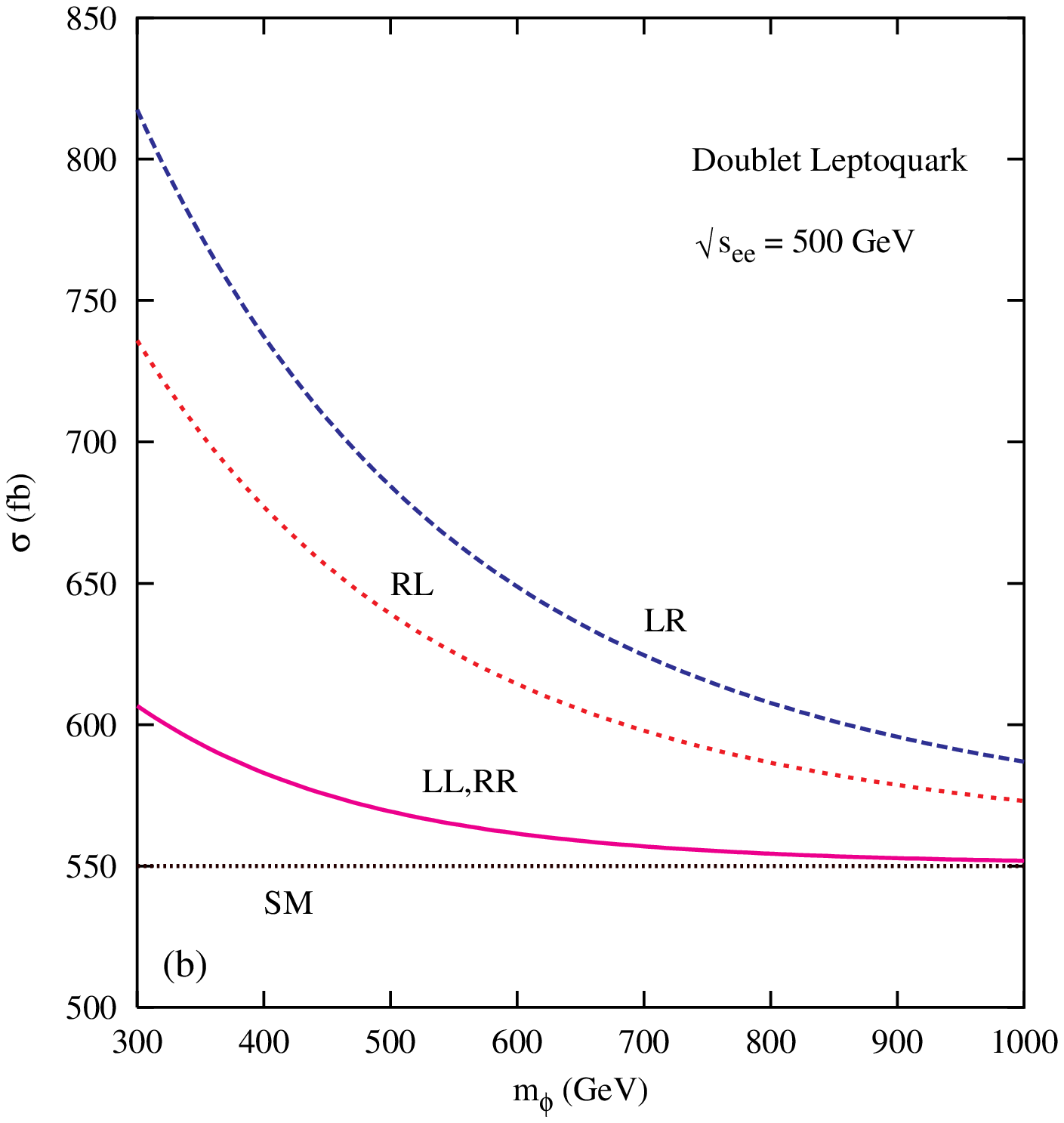}
}

\caption
{{\it Variation of top-pair production cross-section (in presence of (a) 
singlet,
    triplet leptoquarks, R-parity violating interactions and (b)
    doublet leptoquarks) with leptoquark mass ($m_\phi$). For
    comparison we have also plotted the pure SM contribution which is
    independent of $m_\phi$ . $e^+ e^-$ centre-of-mass energy,
    $\sqrt{s_{ee}}$, is fixed at 500 GeV.  Different lines are for different 
 kinds of interactions. Legends follow from the Table 2. The curve 
 marked by $RL$ in (a) corresponds to $R$-parity violating SUSY. }}

\end{figure}


Now let us consider the variation of the cross-section with 
leptoquark mass. For this purpose, we fixed the centre-of-mass energy
of the $e^+e^-$ system at 500 $GeV$. One can easily see that the
leptoquark (or squark) mass acts as the scale of the new physics we
are interested in. This particular feature is reflected in Fig. 3(a)
and (b) where we plotted the total cross-section with $m_\phi$. As
$m_\phi$ increases all the cross-sections are converging to the SM
value, indicating the decoupling nature of the leptoquark interactions
at higher energies. 

\begin{figure}[ht]

\centerline{
\epsfxsize= 7.0 cm\epsfysize=7.0cm
                     \epsfbox{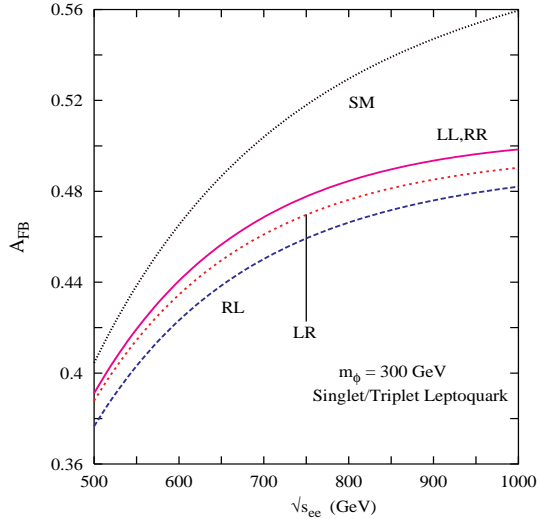}
}

\caption
{{\it Variation of forward-backward asymmetry, $A_{FB}$, (in presence of singlet,
    triplet leptoquarks, R-parity violating interactions) with $e^+ e^-$ 
   centre-of-mass energy . For
    comparison we have also plotted the pure SM contribution.
    Leptoquark mass is fixed at 300 GeV.  Different lines are for different 
 kinds of interactions. Legends follow from the Table 2. The curve 
 marked by $RL$  corresponds to $R$-parity violating SUSY. }}
\label{afb}
\end{figure}

From the above discussions, it is evident that the presence of lepton
number violating couplings may enhance the total rate of top-quark
pair production in electron positron annihilation. Absence of any such
increase in $t \bar t$ cross-section at the future $e^+ e^-$
machines would help us to constrain the parameter space of the
theories which allow such interactions.  As we emphasised, there can
be several types of such interactions. Now it is important to consider
how one can differentiate those if at any experiment such an enhancement
is detected. Different chiral structures of the interactions point to
the fact that angular distribution may be helpful. The most useful
signal of top-quark pair production comes from when one top decays
semi-leptonically and the other decays hadronically. The cleaner
environment of an electron positron collider enables us to reconstruct
the scattering angle from the hadronically decaying top.  So we have
tried to compare the angular distributions of pure SM case with the
leptoquark case. At lower $\sqrt{s} _{ee}$, there is a very little
difference between these cases. At higher centre-of-mass energies
($\sim 1 ~TeV$), angular distribution in leptoquark cases become less
(than the SM) asymmetric in $\cos \theta$ ($\theta$ is the scattering
angle).  To quantify this we calculate the forward backward asymmetry,
$A_{FB}$, defined as,

\be
A_{FB} = \frac{\sigma_B - \sigma_F}{\sigma_B + \sigma_F} 
\ee 
where $\sigma_B = \int^0_{-1}\frac{d\sigma}{d(\cos\theta)} \,
d(\cos\theta)$ and $\sigma_F = \int^1 _0
\frac{d\sigma}{d(\cos\theta)}\, d(\cos\theta)$. 

We have plotted this asymmetry with $e^+ e^-$ centre-of-mass energy in
fig. \ref{afb} for the singlet/triplet leptoquark interactions, along
with the SM. As expected, for the SM, $A_{FB}$ grows with
centre-of-mass energy.  From the figure it is evident, though at lower
energies, $A_{FB}$ for all the four cases remain very close to the SM
value, at higher energies angular distributions for leptoquark
mediated cases become less asymmetric. This in turn reduces $A_{FB}$
in all these cases from the SM value.  Forward-backward asymmetries
for $LL$ and $RR$ cases come out to be equal. At higher energies also,
values of $A_{FB}$, for different kinds of leptoquark interactions
remain very close to each other. So one needs a large number of clean
background free events (which looks possible in the next generation
$e^+e^-$ machines) to differentiate these scenarios.  Once again we
will try to compare our results with that obtained in ref.
\cite{lepto_debchou} in a qualitative manner.  According to this work,
an 1 TeV electron positron collider will explore a larger area in
leptoquark parameter space than a 500 GeV machine. When one looks at
the total cross-sections (see fig. 2a and fig. 2b), one can see that
at higher centre-of-mass energies, differences between the SM
cross-section and that of different leptoquark (+ SM) mediated
processes are less than the differences at lower energies. But when we
look at the forward backward asymmetries at different energies, it is
evident at higher energies the differences between the SM case and
leptoquarks are higher than those evaluated at smaller centre-of-mass
energies. So comparison of the forward backward asymmetry (which is
also the reflection of the angular distribution of the processes) will
be more efficient at higher energies to discriminate the leptoquark
models from the SM which is in consonance with the results in ref.
\cite{lepto_debchou}.

For the doublet leptoquarks, there are no qualitative differences in
$A_{FB}$ from the singlet case. Numerically, for different types of 
doublet leptoquark interactions ($LL, RL, LR$ {\em etc.}) $A_{FB}$
differ very little from the corresponding singlet/triplet cases. We do not
present them here.

Finally we want to make some comments about the top-quark decay
mediated via these new interactions. As we assume this particular
coupling (involving $e$, $t$ and a scalar leptoquark, {\em i.e.}
$\lambda^{\prime}_{132}$, $\lambda^{(i)}_{13}$ or $\tilde
\lambda^{(i)}_{13}$) to be non-zero, top-quark decay width to
$b\,e\,\nu_e$ could also be modified.  We have not written the relevant
interactions involving a $b$-quark, a neutrino and a leptoquark.
Looking at the interactions in ref. \cite{lepto_lag}, one can easily check that
this particular decay cannot be mediated via the $SU(2)$ doublet
leptoquarks. Operators (apart from SM contribution mediated by $W$-boson)
contributing to this process can be written as;
\bea 
Singlet/Triplet &:& |\lambda^{(i)}_{13}|^2 \;\left({\bar e}^c
  \,P_L\, t\right) \;\left({\bar \nu_{e}} \,P_R \,b^c \right)
\widehat{\phi_i \phi^\ast_i}
\nonumber \\
Singlet&: & \lambda^{(1)}_{13} \tilde \lambda^{(1)}_{13}\; \left({\bar
    e}^c \,P_R \,t \right)\;\left({\bar \nu_{e}} \,P_R\, b^c \right)
\widehat{\phi_1 \phi^\ast_1} 
\label{topdecay}
\eea 
\begin{figure}[ht]

\centerline{
\epsfxsize= 7.0 cm\epsfysize=7.0cm
                     \epsfbox{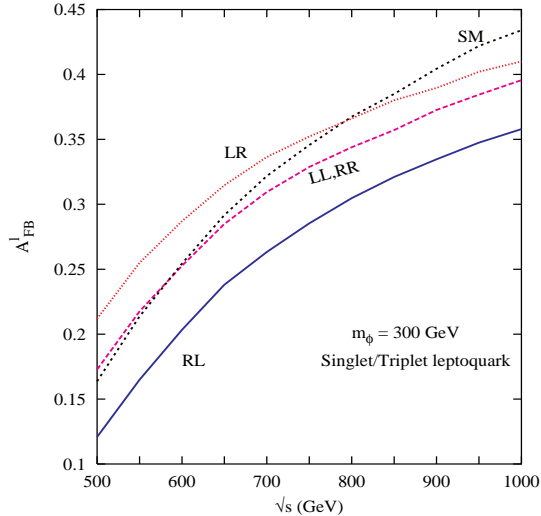}
}

\caption
{{\it Variation of $e^+$ (coming from $t$-quark decay) forward-backward asymmetry, $A^l_{FB}$,
   (in presence of singlet,
    triplet leptoquarks, R-parity violating interactions) with $e^+ e^-$ 
   centre-of-mass energy . For
    comparison we have also plotted the pure SM contribution.
    Leptoquark mass is fixed at 300 GeV.  Different lines are for different 
 kinds of interactions. Legends follow from the Table 2. For top decay 
 we have used the first of eqn. \ref{topdecay}. The curve 
 marked by $RL$  corresponds to $R$-parity violating SUSY. }}
\label{afb_lep}
\end{figure}


R-parity violating SUSY corresponds to the first one of eqn. \ref{topdecay}.
There can be other decay modes, but as long as we confine to
the specific coupling (which we have used so far) this is the only
one.  We have calculated the decay widths corresponding to the cases
in eqn. \ref{topdecay}. With the values of couplings and leptoquark
masses we have used before, the width comes out to be very nearly
equal to the SM value. This looks surprising because with the same
values of the parameters we get pretty good enhancement in $t \bar t$
production. The smallness of new-physics contribution can be
attributed to the fact that, dominant contribution to the amplitudes,
corresponds to eqns.  \ref{topdecay}, are proportional to $m_t \,m_b$
when in the case of the top-pair production these are proportional to
$m_t^2$. So the top semi-leptonic branching ratio (to electron) is
barely changed in the presence of these new interactions, unless the
couplings are big enough. 

Operators responsible for top-quark decay (eqn. \ref{topdecay}) have a
distinctly different structure form the SM case. Though the total
width shows a little enhancement over the SM value, it would be
interesting to see how the angular distribution of the decay products
differ from the later. As we pointed out, the cleanest signal for top
pair production comes from when one top decays semileptonically and
the other decays hadronically. We have calculated the angular
distribution of the $e^+$ coming from the {\em top} decay keeping the
full spin correlation between the top production and decay, in
presence of leptoquark interactions as well as the SM. From the
angular distribution one can easily calculate forward backward
asymmetry of the $e^+$ ($A^l_{FB}$).  For the purpose of illustration, we
have presented the result of our analysis for singlet/triplet
leptoquarks in fig \ref{afb_lep}.  We have chosen the first one of
eqn. \ref{topdecay} to calculate the top-quark decay matrix element.
Fig \ref{afb_lep} clearly shows the difference in $A^l_{FB}$ between
the SM and leptoquarks interaction over the energy range we have
considered.  Despite of the fact that these new interactions (with the
coupling strength we have considered) could not change the top
semileptonic branching ratio to a significant extent, angular
asymmetries still play a crucial role in discriminating these effects
from the SM. With the ballpark values of the $t \bar t$ cross-sections
at these energies (see fig. 2a) and with the projected $e^+e^-$
luminosities, one can easily detect these asymmetries. A comparison of
fig. \ref{afb_lep} with fig. \ref{afb}, reveals that the $A_{FB}$ in
$t\bar t$ production differs from that of $A^l_{FB}$ over the whole
range of centre-of-mass energy. This can be accounted by the chiral
structure of decay matrix element which plays a crucial role in determining
the angular distribution of the top decay products.

R-parity violating MSSM allows the $t$-quark to decay to left-handed
selectron ($\tilde e_L$) and a $b$-quark via the same $\lambda^\prime
_{132}$ coupling.  $\tilde e_L$ will in turn decay to a electron and
to the lightest neutralino ($\tilde \chi^0 _1$). $\tilde \chi^0 _1$ is
no longer stable and would decay to $s$-quark, $\nu_e$ and $b$-quark.
This has been discussed in detail in ref. \cite{t_dk_agashe}. This
decay will lead to 3 jets (including one $b$-quark), an electron and
missing energy originating from a neutrino. So R-parity violation can
be separated out from non-SUSY leptoquarks by this kind of top-decay
signals.

\section{Conclusion}
To summarise, we show that presence of lepton number violating
interactions can enhance the top-quark pair production cross-section
in electron positron annihilation at next generation linear collider
machines. We have considered different kinds of leptoquark
interactions. R-parity violating interactions, involving one lepton,
and two quark superfields, belong to one of these above cases.
Non-SUSY theories with leptoquarks allow both left and right handed
couplings involving a scalar leptoquark, a top-quark and an electron.
We have estimated the cross-sections in all the cases separately. With
moderate values of these lepton number violating Yukawa couplings one
gets pretty good enhancement of total cross-section over the SM value.
Depending on the $e^+ e^-$ centre-of-mass energy and leptoquark mass,
enhancement varies from a few percent to 60$\%$. With higher values of
leptoquark mass cross-section converges to the SM value. This clearly
points to the fact that these interactions are decoupling in nature at
higher energies. We have also considered the effects of this coupling
on the top semi leptonic decay.  Top-decay width changes very little
after inclusion of these new interactions. Forward-backward
asymmetry in top pair production and top-decay may be used to
differentiate these lepton number violating interactions from
the SM and among themselves at higher centre-of-mass energies. But
this will need a large sample of $t \bar t$ events which looks feasible
at the next generation $e^+e^-$ linear colliders.

{\bf Acknowledgement}

The author thanks A. Raychaudhuri for carefully reading the manuscript
and P. Konar for computational help. Thanks are also due for D.
Choudhury for pointing me out some earlier important works in this
direction and for many useful suggestions regarding this work.

\newcommand{\plb}[3]{{Phys. Lett.} {\bf B#1} #2 (#3)}                  %
\newcommand{\prl}[3]{Phys. Rev. Lett. {\bf #1} #2 (#3)}        %
\newcommand{\rmp}[3]{Rev. Mod.  Phys. {\bf #1} #2 (#3)}             %
\newcommand{\prep}[3]{Phys. Rep. {\bf #1} #2 (#3)}                     %
\newcommand{\ijmp}[3]{Int. J. Mod. Phys. {\bf A#1} #2 (#3)} 
\newcommand{\rpp}[3]{Rep. Prog. Phys. {\bf #1} #2 (#3)}             %
\newcommand{\prd}[3]{{Phys. Rev.}{\bf D#1} #2 (#3)}                    %
\newcommand{\np}[3]{Nucl. Phys. {\bf B#1} #2 (#3)}                     %
\newcommand{\npbps}[3]{Nucl. Phys. B (Proc. Suppl.) 
           {\bf #1} (#3) #2} %
\newcommand{\sci}[3]{Science {\bf #1} #2 (#3)}                 %
\newcommand{\zp}[3]{Z.~Phys. C{\bf#1} #2 (#3)}                 %
\newcommand{\mpla}[3]{Mod. Phys. Lett. {\bf A#1} #2 (#3)}             %

\newcommand{\astropp}[3]{Astropart. Phys. {\bf #1} #2 (#3)}            %
\newcommand{\ib}[3]{{\em ibid.\/} {\bf #1} #2 (#3)}                    %
\newcommand{\nat}[3]{Nature (London) {\bf #1} (#3) #2}         %
\newcommand{\nuovocim}[3]{Nuovo Cim. {\bf #1} #2 (#3)}         %
\newcommand{\yadfiz}[4]{Yad. Fiz. {\bf #1} (#3) #2 [English            %
        transl.: Sov. J. Nucl.  Phys. {\bf #1} #3 (#4)]}               %
\newcommand{\philt}[3]{Phil. Trans. Roy. Soc. London A {\bf #1} #2  
        (#3)}                                                          %
\newcommand{\hepph}[1]{hep--ph/#1}           %
\newcommand{\hepex}[1]{hep--ex/#1}           %
\newcommand{\astro}[1]{(electronic archive:     astro--ph/#1)}         %
\newcommand{\epj}[3]{Euro. Phys. J {\bf C#1} #2 (#3)}  


\end{document}